%% file: main.tex
\documentstyle[12pt]{article}
\begin{document}
\def\D#1{{\bf D#1}}
\def\B#1{{\bf B#1}}
\def\C#1{{\bf C#1}}
\input epsf.tex
\input title.tex

\vskip 0.8cm
%
\input sumary.tex
\vskip 0.8cm
\topmargin -1cm
{\small
\input chap1.tex

\vskip 0.8cm
%
\input chap2.tex

%
\vskip 0.8cm
%
\input how.tex
%
\vskip 0.8cm
\input remark.tex
%
%
\vskip 0.8cm
\input ack.tex
\vskip 0.8cm
\input ref.tex
\newpage
\vskip 0.8cm
\vskip 0.8cm
%
\input appa.tex
\vskip 0.6cm
%
\input tbl.tex
%
\vskip 0.8cm
%
\input install.tex

\vskip 0.8cm
\input testrun.tex

}
\end{document}

%% file: title.tex
\begin{flushright}
CP--043
\end{flushright}
\par
\vskip 2cm
\centerline{\large {\Large {\tt grc4f}} v1.0: 
a Four-fermion Event Generator}
\centerline{\large for $e^+e^-$ Collisions }
\par
\vskip 0.3cm
{\small
\centerline{J. Fujimoto, T. Ishikawa, T. Kaneko$^{a,d)}$, K. Kato$^{b)}$,}
\centerline{S. Kawabata, Y. Kurihara, T. Munehisa$^{c)}$,}
\centerline{D. Perret-Gallix$^{d)}$, Y. Shimizu and H. Tanaka$^{e)}$} 
}
\vskip 0.2cm
{\footnotesize\it
\centerline{KEK, Oho, Tsukuba, Ibaraki 305, Japan}
\centerline{a) Meiji-Gakuin University, Kamikurata, Totsuka 244, Japan}
\centerline{b) Kogakuin University, Nishi-Shinjuku, Tokyo 160, Japan}
\centerline{c) Yamanashi University, Takeda, Kofu 400, Japan}
\centerline{d) LAPP, B.P. 110, Annecy-le-Vieux F-74941 Cedex, France}
\centerline{e) Rikkyo University, Nishi-Ikebukuro, Tokyo 171, Japan}
}
\par
\vskip 0.3cm
%
\vskip 0.2cm
\noindent
\unitlength 1mm
\begin{picture}(160,1)
\put(0,0){\line(1,0){160}}
\end{picture}
\par
%
\leftline{\small\bf Abstract}
{\footnotesize
{\large {\tt grc4f}} is a Monte-Carlo package
for generating $e^+e^-\to 4\hbox{-fermion}$ processes in 
the standard model. All of the 76 LEP-2
allowed fermionic final state 
processes evaluated at tree level are included in version 1.0.
{\tt grc4f} addresses event simulation requirements at $e^+e^-$
colliders such as LEP and up-coming linear colliders.
Most of the attractive aspects of {\tt grc4f} come from its link
to the {\tt GRACE} system: a Feynman diagram automatic computation 
system.
The {\tt GRACE} system has 
been used to produce the computational code for all final states, giving a
higher level of confidence in the calculation correctness. 
Based on the helicity amplitude calculation technique, 
all fermion masses can be kept finite
and helicity information can be propagated down to the final state particles.
The phase space integration of the matrix element 
gives the total and differential cross sections, then unweighted events
are Generated. Initial state 
radiation (ISR) corrections are implemented in two ways, one is based on the
electron structure function 
formalism and the second uses the parton shower algorithm called {\tt QEDPS}. 
The latter can also
be applied for final state radiation (FSR) though the interference with the
ISR is not yet taken into account. Parton shower and hadronization of 
the final quarks are performed through an interface to 
{\tt JETSET}. Coulomb correction between two intermediate $W$'s, 
anomalous coupling as well as gluon contributions
in the hadronic processes are also included. 
}
\par\noindent
\begin{picture}(160,1)
\put(0,0){\line(1,0){160}}
\end{picture}
\par

%% file: sumary.tex
%
\noindent
\begin{minipage}{7.6cm}
\leftline{\small\bf PROGRAM SUMMARY}
\medskip
{\footnotesize
\noindent
{\it Title of program}: {\tt grc4f}
\par
\noindent
{\it Program obtainable from}:~CPC Program Library, Queen's University
of Belfast, N.Ire\-land ( see application form in this issue ) and from
ftp.kek.jp in directory kek/minami/grc4f.
\par
\noindent
{\it Computer for which the program is designed and others on which it is 
operable}:~HP9000 and most of the UNIX platforms with a FORTRAN77 compiler 
\par
\noindent
{\it Computer}: HP9000; ~{\it Installation}: National Laboratory for
High Energy Physics (KEK), Tsuku\-ba, Ibaraki, Japan
\par
\noindent
{\it Operating system}:~UNIX
\noindent
{\it Programming language used}: FORTRAN77
\par
\noindent
{\it High speed storage required}: 36 Mbyte 
\noindent
{\it Card image code}: ASCII
\par
\par
}
\end{minipage}
~~~
\begin{minipage}{7.6cm}
\medskip
{\footnotesize
\noindent
{\it Nature of physical problem}
\par
\noindent
Study of $W$ boson physics and of background processes
for new particle search at LEP-2 and beyond.
\par
\noindent
{\it Method of solution}
\par
\noindent
The automatic amplitude generator {\tt GRACE} is used to get the
necessary helicity amplitudes for all the four-fermion processes.
Specific corrections are then introduced to deal with 
the gauge boson width, the radiative corrections, the color assignment, 
the hadronization, the coulomb correction and the anomalous coupling.
\par
\noindent
{\it Typical running time}
\par
\noindent
The running time depends on the number of diagrams of the
selected process, on the required cross-section accuracy and
on the applied cuts. 
For instance, the process $e^-\bar{\nu_e}u\bar{d}$ takes 20
minutes to reach a 0.5\% accuracy on the total cross section integration 
with HP-735/99.
\par
\noindent
}
\end{minipage}

%% file: chap1.tex
\medskip
\leftline{\large\bf 1. Introduction}
\smallskip

The main purpose of the LEP-2 experiments is to measure the properties
of the $W$-boson with a high level of precision by its direct pair production.
At the $Z^0$ peak (LEP-1 energy),
two-fermion final states produced by the $Z^0$ decay are, by far, the
most important processes. At higher energy, charged $W$-boson being 
necessarily created
in pair at $e^+e^-$ colliders, the cross-section is dominated
by four-fermion processes.
However, other production mechanisms may also lead to four-fermion 
final states like $Z^0$ pair or
two-photon like production. 
In addition, some four-fermion processes contribute heavily to the background
for new particle searches due to their large missing energy. 
In this paper, a four-fermion event generator, {\tt grc4f}, is presented.
It provides a convenient way of computing cross-sections under the
complex cuts, acceptance and resolution of the collider experiments.
 
It is based on the {\tt GRACE}\cite{grace} system which generates 
automatically the matrix element in terms of helicity amplitudes 
(supplied by the {\tt CHANEL}\cite{chanel} library) for 
any process once the initial and the final states have been specified. 
At present,
the system has been completed only at tree level in the framework 
of the standard model for electroweak and strong interactions. 
Fermion masses are non-zero and helicity information can be traced
down to the final state particles.
In addition, 
a kinematics library has been developed to cover the requirements
of each process topology 
(weeding out multidimensional singularities) to get a faster 
convergence of the Monte-Carlo integration over the phase 
space. 
The {\tt grc4f} package is actually a collection of 76
$e^+e^-\to4\hbox{-fermion}$ processes 
\footnote{In version v1.0,
final states with top quarks or those obtained by complex conjugation are
not explicitly included.
For generating $t$-quark final state processes,
the user may try
to change the $c$-quark mass to the top mass 
together with an interchange $s \leftrightarrow b$,
but the integration convergence
is not warranted because $t\rightarrow bW$ is a real process contrary
to $c \rightarrow sW$.
}, 
consisting of 16 hadronic, 36 semi-hadronic and 24 leptonic
processes, presented in a coherent and 
uniform environment. 
Flavor mixing is set to zero.
The coupling between electron and Higgs boson is suppressed. 
Once a process has been selected, the total and differential cross 
sections are computed
with the Monte-Carlo integration package {\tt BASES} \cite{bases}.
Then {\tt SPRING} \cite{bases}, a general purpose event generator, 
provides unweighted events.

The matrix element generated by {\tt GRACE} corresponds
to the genuine tree level process. 
However, several corrections must be introduced to produce
realistic cross sections.
First of all the most basic correction, inherent to heavy boson 
production, is due to the finite width of the $W$ or
$Z^0$ boson to ensure a finite cross section.
It should be noted that if one introduces the width
in a naive way, it violates the gauge invariance 
and in some case may lead to a divergent cross-section. 
The $e^+e^- \rightarrow e^-{\bar \nu_e}q{\bar q'}$ cross section, for example,
blows up at $\theta_e\approx 0$. To cure this problem, the terms
which satisfy the Ward identity are properly subtracted from the electron
current which is connected to the $t$-channel photon\cite{perretg}.
The computation of the 
constant and the running decay widths of $W^\pm$ and $Z^0$ is presented
in the section 2.1.

Two additional effects, although of different origins, can be, 
in first approximation, introduced into any
selected process; radiative corrections
and hadronization of the final quarks.  

For the radiative corrections two techniques are proposed
in the program.
The first one uses the well-known analytic form of $e^\pm$ structure 
function\cite{kuraev} 
and the second is based on {\tt QEDPS}\cite{isr}, a radiative correction
generator
producing an indefinite numbers of photons according to the parton 
shower algorithm in the leading-logarithmic (LL) approximation. 
Originally this
algorithm has been developed to simulate QCD parton shower.
One important point here is that {\tt QEDPS} reproduces
the radiative photon transverse momentum distributions. The details of
these two methods shall be given in section 2.2.
Final state radiation (FSR) for the electron and muon 
can also be generated with {\tt QEDPS}.
However the contribution of the interference between ISR and FSR is 
{\sl not} yet included. 
Hence one should not use the FSR mode 
when the final state contains electron for kinematical regions where the $t$-channel
photon exchange is the dominant diagram. The FSR for the quarks is not
allowed in {\tt grc4f}, because the QCD evolution scale is much shorter
than the QED one, therefore the photonic correction on quarks might be 
meaningless.
We recommend to use the FSR mode only for the case where
the $W$-pair or $Z$-pair ($Z\gamma$) production diagrams are dominant.

We assume that the hadronization of partons can be separated
from the hard interaction studied here.
Under this assumption, the calculation of cross sections is
exact in {\tt grc4f}. 
Final state parton hadronization is performed, in {\tt grc4f},
through the {\tt LUND} mechanism as implemented in 
{\tt JETSET}\cite{jetset}. A set of color bases is defined to
select final state color flow. Possible ambiguities in this
procedure as well as other QCD related issues are discussed in
details in section 2.3.
 
In order not to waste cpu time and disk space in the generation of events
that will be eventually rejected by detector acceptance or threshold,
a set of general cuts whose values can be set by the user have been
implemented directly in the kinematics as presented in section 2.4.

The Coulomb correction\cite{coulomb} due to the exchange of a virtual photon
between two intermediate $W$ 
is important for non-relativistic $W$, namely close to the production
threshold. It may reach 4\% of the total cross-section. The introduction
of this effect is described in section 2.5.

Anomalous couplings in the interaction among vector bosons is introduced
in section 2.6.

The basic input 
parameters used in the program are listed hereafter. 
Fermion masses 
are taken from the 
report of Particle Data Group~\cite{PDG}, but can be set to zero if 
necessary.
\begin{eqnarray*}
        M_W&=&   80.23         \hbox{ GeV},  \nonumber\\
        M_Z&=&      91.1888     \hbox{ GeV},   \nonumber\\
        \Gamma_W&=&3G_FM_W^3/(\sqrt{8} \pi), \qquad
        G_F=1.16639 \cdot 10^{-5} \hbox{ GeV}^{-2},\nonumber \\
        \Gamma_Z&=&   2.497 \hbox{ GeV},   \nonumber\\
        \alpha(0)&=&  1/137.036,            \qquad
        \alpha(M_W)=     1/128.07,        \nonumber\\
        \alpha_S&=&     0.12.         \nonumber
\end{eqnarray*}
Here $\alpha(Q)$ is the QED fine structure constant at the energy 
scale $Q$ and $\alpha_S$ is the QCD coupling. 
These values can be changed
through control data. The weak mixing angle is calculated
by $\sin^2\theta_W=1-M_W^2/M_Z^2$.

The matrix elements are given in term of helicity amplitudes, 
it is therefore possible to select any helicity state configuration
(initial and final). 

This paper is organized as follows. Theoretical details for the
problems mentioned in the introduction will be given in 
section~2. The structure of the program is described in 
section~3. 
In section 4, all details about
running the program are presented. Some improvement
for future releases are summarized in the section 5. 
Three appendices present the parameters and options 
which can be changed by the user, the list of all processes and the 
program installation procedure followed by a run example. 

\medskip
\leftline{\large\bf 2. Theoretical aspects} 
\smallskip
This section 
covers the treatment of the boson 
width, the introduction of the initial and final state radiation, the 
QCD related
issues such as hadronization and the color problems, the description of 
kinematics and cuts, the Coulomb correction and the anomalous
coupling of heavy vector bosons.

\medskip
\leftline{\large\bf 2.1 Gauge boson width}
\smallskip

The $O(\alpha)$ self-energy of the gauge boson,
$\Sigma^{(1)}(k^2)$, generally satisfies
\begin{eqnarray*}
{\rm Re}\Sigma^{(1)}(M^2)&=&0, \nonumber \\
{\rm Im}\Sigma^{(1)}(M^2)&=&-M\Gamma(M^2),
\end{eqnarray*}
where $M$ is the mass of the gauge boson and $\Gamma(M^2)$ is 
the lowest order decay width.
This induces the replacement of the gauge boson propagator 
by the following simple Breit-Wigner form:
\begin{eqnarray*}
{1\over k^2-M^2} \rightarrow {1 \over k^2-M^2+iM\Gamma(M^2)}
\end{eqnarray*}
This form is used for the fixed width scheme in {\tt grc4f} whenever 
the propagator appears with positive momentum squared.

Another way is to take into account the energy dependence of
the self-energy. Based on the observation
that the contribution of fermion pairs to {\rm Im}$\Sigma^{(1)}$
is $k^2$ times a constant in any gauge as long as all the light 
fermion masses are neglected,
the following energy dependent width has been
proposed~\cite{width}.
\begin{eqnarray*}
M\Gamma(M^2) \rightarrow -{\rm Im}\Sigma^{(1)}(k^2)|_{{\rm
   fermion}~{\rm pairs}} \cong {k^2 \over M}\Gamma(M^2).
\end{eqnarray*}
In {\tt grc4f} this form is used as the default, but
the constant width can be selected as well.

As mentioned in the previous section, the introduction of the boson
width may raise difficulties, particularly when the electron is
scattered in the very forward direction. 
Such final states usually come from the generation
of the so-called single $W$ or $Z$ process.
To get rid of the divergent cross section, we apply the following
method.

With an arbitrary gauge parameter $\xi$ the electron current can be written as:
\begin{eqnarray}
 l_\mu &=&{\bar u}(p')\gamma^\nu u(p)
[g_{\mu \nu}+(\xi -1) k_\mu k_\nu/k^2],\\
            &=&{\bar u}(p')\gamma_\mu u(p),
\end{eqnarray}
where $p_\mu$($p'_\mu$) is the four-momentum of the initial (final)
electron and $k_\mu=p_\mu-p'_\mu$, the momentum of the virtual
photon.
After squaring the amplitude and averaging over spin states, 
the matrix element can be written as

\begin{equation}
   M = L_{\mu\nu}T^{\mu\nu},
\end{equation}
with
\begin{equation}
L_{\mu \nu}=\overline{\sum_{spin}} l_\mu l_\nu^*
=2\left[p_\mu p_\nu' + p_\nu p_\mu'+\frac{k^2}{2} g_{\mu \nu}\right].
\end{equation}
Let's assume that $T_{\mu\nu}$ is gauge invariant.
Then one can replace $L_{\mu \nu}$ by
\begin{eqnarray}
L_{\mu \nu}  \to L_{\mu \nu}' = 4 p_\mu p_\nu +  k^2  g_{\mu \nu}.
\end{eqnarray}
In this form the first term is responsible for the blow-up of the cross
section. Thanks to the gauge invariance,
one can further replace the vector $p_\mu$ by:
\begin{eqnarray}
p_\mu \to P_\mu = p_\mu-(p_0/k_0)k_\mu.
\end{eqnarray}
By substituting $P_{\mu}$ in Eq.(5) and
dropping $k_\mu$, one gets the final form
\begin{eqnarray}
L_{\mu \nu}'  \to L_{\mu \nu}'' = 4 P_\mu P_\nu + k^2 g_{\mu \nu}.
\end{eqnarray}
It is known that a product of $P$ with an arbitrary vector $A$,
$P \cdot A$, can be expressed by a sum of terms
proportional to either $m_e^2$, $1-\cos{\theta_e}$
or $\sin{\theta_e}$. The part $T_{\mu\nu}$ is expressed in terms of some
momenta which are to be contracted with $P_\mu$ or $P_\nu$.
Hence in the region $\theta_e \approx 0$, any product
behaves effectively like $k^2$, because $1-\cos{\theta_e}$ vanishes
almost like $k^2$. If the current
$L_{\mu \nu}''$ is used instead of the original one $L_{\mu \nu}$,
the cross-section remains finite down to  $\theta_e = 0$.

\medskip
\leftline{\large\bf 2.2 Radiative corrections}
\smallskip
In the first approach, the simple electron structure function 
is used.  
The electron structure function at
$O(\alpha^2)$~\cite{kuraev}  which is to be
convoluted to the cross section for a primary
process is given by
\begin{eqnarray}
  D(x,s)&=& \left[1+{3\over8}\beta
 +\left({9\over128}-{\zeta(2)\over8}\right)\beta^2\right]
                 {\beta\over2}(1-x)^{\beta/2-1} \nonumber \\
    && -{\beta\over 4} \bigl( 1+x \bigr) 
        -{\beta^2 \over 32} \biggl[ 4(1+x)\ln(1-x) 
              +{1+3x^2 \over 1-x}\ln x+(5+x) \biggr], 
                    \label{eq:Dxs} \\
\beta&=&(2\alpha/\pi)(\ln(s/m_e^2)-1),
\end{eqnarray}
where $s$ is the square of the total energy of the system and $x$ is 
the momentum fraction of the electron.
It should be noted that, when 
compared with the exact $O(\alpha)$ calculation,
the radiative corrected cross section obtained by this function
does not contain the overall multiplying $K$-factor: 
\begin{equation}
1+{\alpha\over\pi}\left({\pi^2\over3}-{1\over2}\right)=1.006480\cdots.
\label{eq:kfact}
\end{equation}
In {\tt grc4f} this factor is missing both for the structure function
mode and {\tt QEDPS} mode. 
If necessary, the final result can be multiplied by this factor for better
accuracy.

The QED Parton Shower approach, {\tt QEDPS}, 
is primarily based on the fact that $D(x,Q^2)$ obeys the 
Altarelli-Parisi equation:
\begin{equation}
{d D(x,Q^2)\over d\ln Q^2}={\alpha\over 2\pi}
                               \int_x^1 {dy \over y} P_+(x/y) D(y,Q^2),
                                        \label{eq:AP}
\end{equation}
in the leading-log(LL) approximation~\cite{ll}. 
This is equivalent to the following integral equation:
\begin{equation}
D(x,Q^2)= \Pi(Q^2,Q_s^2)D(x,Q_s^2)
+{\alpha\over2\pi}\int\nolimits_{Q_s^2}^{Q^2}{dK^2\over K^2}
    \Pi(Q^2,K^2)\int\nolimits_x^{1-\epsilon}{dy\over y}
           P(y)D(x/y,K^2),       \label{eq:intform}
\end{equation}
where the small quantity $\epsilon$ will be specified later.
In these equations $P(x)$ is the split function and  $P_+(x)$ 
is that with regularization at $x=1$.
$Q_s^2$ is the initial value of $Q^2$. 
For simplicity the fine structure constant $\alpha$ is assumed not 
running with $Q^2$. 
The Sudakov factor $\Pi$ is given by:
\begin{equation}
 \Pi(Q^2,{Q'}^2) = \exp\left(- {\alpha\over 2 \pi} \int_{{Q'}^2}^{Q^2}
 { d K^2 \over K^2} \int_0^{1-\epsilon} d x P(x) \right).
  \label{eq:non}
\end{equation}
This is the probability that an electron evolves from ${Q'}^2$ to 
$Q^2$ without emitting hard photon. In other words, $\Pi$ already 
contains the soft photon component, which causes 
the change in the electron virtuality, and the loop correction 
contribution at all orders of perturbation.

The integral form Eq.(\ref{eq:intform}) can be solved by iteration.
It is clear that 
the emission of $n$ photons corresponds to $n$ iterations. Hence 
it is possible to regard the process as a stochastic mechanism suggesting 
the following shower algorithm~\cite{ps1}.
\begin{description}
\item[(a)] Set $x_b=1$. The variable $x_b$ is the fraction of the
    light-cone momentum of the virtual electron that annihilates.
\item[(b)] Choose a random number $\eta$. If it is smaller than 
    $\Pi(Q^2,Q_s^2)$, then the evolution stops. If not, one can find 
    the virtuality $K^2$ that satisfies $\eta =\Pi(K^2,Q_s^2)$ at 
    which a branching takes place.
\item[(c)] Fix $x$ according to the probability $P(x)$ between 0 and 
    $1-\epsilon$. Then $x_b$ is replaced by $x_b x$.
    Go back to (b) by substituting $K^2$ into $Q_s^2$ and 
    repeat until the evolution stops.
\end{description}
Once a radiative process is fixed by this algorithm, each branching of
a photon is a real process, that is, 
an electron with $x,K^2$ decays like:
\begin{equation} 
      e^-(x,-K^2)\to e^-(xy,-{K'}^2)+\gamma(x(1-y),Q_0^2).
 \label{eq:elsplit}
\end{equation}
Here we have introduced a 
cutoff $Q_0^2$ to avoid the infrared divergence. The momentum
conservation at the branching gives:
$ -K^2=-{K'}^2/y+Q_0^2/(1-y)+{\bf k}_T^2/(y(1-y)) $
which in turn determines the photon transverse momentum relative to
the parent, ${\bf k}_T^2$, from $y,K^2,{K'}^2$. 
This technique gives the ${\bf k}_T^2$ distribution as well as
the shape of $x$.

The kinematical boundary $y(K^2+Q_0^2/(1-y))\le{K'}^2$,
equivalent to ${\bf k}_T^2>0$, fixes $\epsilon$ as
$ \epsilon=Q_0^2/{K'}^2 $
since $K^2\ll {K'}^2$ is expected.
In ref.\cite{ps3} the important role played by
this $\epsilon$ is discussed in more details.
 
The above description of the algorithm concerns the 
case where either $e^-$ or $e^+$ radiates photons 
when the axial gauge vector is chosen along the momentum of the other 
electron, namely $e^+$ or $e^-$. In the program, however, we 
use the double cascade scheme to 
ensure the symmetry of the radiation between $e^+$ and 
$e^-$~\cite{double}. These two are mathematically
equivalent in the LL approximation. 

The two parameters $Q_s^2$ and $Q_0^2$ are given as follows;
\begin{equation}
    Q_s^2=m_e^2e=m_e^2\times2.71828\cdots,\qquad 
    Q_0^2=10^{-12}~~\hbox{GeV}^2.
\end{equation}
$Q_s^2$ is defined so as to include
the constant term $-1$ of $\beta$ in such a way that
$\beta=(2\alpha/\pi)(\ln(s/m_e^2)-1)=(2\alpha/\pi)\ln(s/(m_e^2e))$.
Since $Q_0^2$ is unphysical, any observable 
should not depend on it. It has been checked that increasing
$Q_0^2$ up to $O(m_e^2/10)$ leaves the result unchanged within the
statistical error of the event generation~\cite{isr}.

This scheme can be applied for the radiation from the final state charged particles as well.
The lower bound of the virtuality integration, $Q_s^2$, is now the
mass square of the final particle instead of the initial electron mass. 
The upper bound is the four-momentum squared of the lepton pair. 
Here we assume that the lepton-pairs are created from the gauge boson.
The FSR should be used for the
processes in which the $W$-pair or $Z$-pair ($Z\gamma$) production diagrams
are dominating, but not for the processes such as multi-peripheral
two-photon like 
diagrams.  

\medskip
\leftline{\large\bf 2.3 QCD related issues}
\smallskip
\medskip
\leftline{\large\bf 2.3.1 Color bases}
\smallskip


We calculate the matrix element using {\tt GRACE} which may be written
as
\begin{equation}
M=| \sum T_j |^2.
\end{equation}
Each amplitude \(T_j\) includes a color factor.
The color indices of the external particle must be
summed in the final state and averaged
in the initial state.

If a diagram has four gluon 
vertices, the diagram is separated into
three pieces, the so-called $s$-, $t$- and
$u$-channels, respectively, for each vertex.
Let's now consider the $T_j$'s not as amplitude but
as a component of an amplitude after this decomposition.

For each $T_j$, the color factor can be factorized out and
expanded on a set of color bases, \(C_k\):
\begin{equation}
T_j = \left(\sum_k w_j^{(k)}C_k \right)\tilde{T}_j
\end{equation}
where \(w_j^{(k)}\)'s are numbers.  
Then the matrix element is given by
\begin{equation}
M= \sum_{k,k'} \left( C_k C_{k'}^{\dagger}\right) F_k F_{k'}^{*}
, \qquad F_k=\sum_j w_j^{(k)} \tilde{T}_j\ .
\end{equation}

Therefore, for the four-gluon vertices,
the colored part of an amplitude 
consists of three-gluon vertices $(-if_{abc})$,
quark-gluon vertices $(t^a)$
and non-colored vertices. 
The ghost-gluon vertex bears the same color structure
as the three-gluon vertex and do not require a specific treatment.

The choice of the color bases and the technique used to introduce color
flow at the event generation level are not all equivalent.
Let's describe one of the simplest approach. The following algorithm 
has been introduced in {\tt GRACE}.
\begin{enumerate} \itemsep=0pt
\item We consider a process in which $n_q$ quark pairs and $n_g$
gluons exist among external particles. 
For any pair of amplitudes, \(T_j\) and
\(T_k^{\dagger}\), the color index of an external
gluon is contracted between them. Making use of this,
each gluon is converted into a pair of quark and antiquark 
creating a quark-gluon vertex. An overall factor \(2^{n_g}\) 
is assigned to \(M\) from:
\begin{equation}
 \delta^{ab}={\rm Tr\;} t^a t^b /T_R, \qquad T_R=\frac{1}{2}.
\end{equation}
\item 
External particles are
\(n\) pairs of quarks and antiquarks where \(n=n_q+n_g\).
Quarks and antiquarks are denoted as \(1, 2, 3, \ldots , n\) and
\(\bar{1}, \bar{2}, \bar{3}, \ldots , \bar{n}\), respectively.
The color base here has the form: 
\begin{equation}
C_k=\delta_{1j_1}\delta_{2j_2} \cdots \delta_{nj_n}
\label{eq:cbase}
\end{equation}
where \(j_1, j_2, \ldots , j_n\) is a permutation of
\(\bar{1}, \bar{2}, \bar{3}, \ldots , \bar{n}\). 
The number of color bases is hence \(n!\).
\item Each three-gluon vertex is converted into
a pair of quark loops by 
\begin{equation}
 -if_{abc}= (-{\rm Tr\;} t^at^bt^c + {\rm Tr\;} t^bt^at^c )/T_R.
\end{equation}
\item Each gluon propagator is replaced by quark lines by use of
the Fiertz transformation:
\begin{equation}
(t^a)_{ij}\delta^{ab}(t^b)_{kl}=
-\frac{1}{2N_C}\delta_{ij}\delta_{kl}
+\frac{1}{2}\delta_{il}\delta_{jk},
\quad (N_C=3)
\end{equation}
\item After the above procedures, only quark lines remain.
For each closed quark loop, a factor \({\rm Tr\;}1=3\) is
assigned.
\end{enumerate}

When the calculation
is done in the covariant gauge, the diagrams including external
ghost particles, if any, are to be considered 
separately\footnote{This does not happen in {\tt grc4f}.}.

A different approach can be followed without gluon-quark 
conversion. For example,
let us denote quarks, antiquarks, and gluons 
as \(1, 2, 3, \ldots , n_q\), 
\(\bar{1}, \bar{2}, \bar{3}, \ldots , \bar{n}_q\), and
\(\hat{1}, \hat{2}, \hat{3}, \ldots , \hat{n}_g\), 
respectively. 
Instead of Eq.(\ref{eq:cbase}), the color bases can be represented by 
a product of the following objects, 
\[
 \delta_{1j_1},\ \
 (t^{\hat{k_1}})_{1j_1},\ \
 (t^{\hat{k_1}}t^{\hat{k_2}})_{1j_1},\ \ \ldots
\]
where \(j_1, j_2, \ldots \) and
\(k_1, k_2, \ldots \) are permutation of
\(\bar{1}, \bar{2}, \bar{3}, \ldots , \bar{n}_q\) and
\(\hat{1}, \hat{2}, \hat{3}, \ldots , \hat{n}_g\),
respectively. 

\medskip
\leftline{\large\bf 2.3.2 Interface to hadronization}
\smallskip
%
As mentioned in the previous section, there are
many possible choice for selecting a color base.
As far as the total cross section is concerned they give the same
result.
However, the choice of the color base and the technique used to introduce
color flow at the generation level are not equivalent and may induce
noticeable discrepancies after hadronization.

We take the string picture used in {\tt JETSET}
as a concrete example of hadronization.
The color flow pattern can be assigned to each
color base intuitively. However, an event
is generated by matrix element which is the square
of the amplitude, i.e., a linear combination of
color bases. For interference terms,
assignment of a color flow pattern is ambiguous. One possible solution
is to introduce a set of orthogonal bases.

To illustrate this point, we consider the
four-quark final state, \(q_1\bar{q_2}q_3\bar{q_4}\)
 which is produced by
{\tt grc4f}. We label them 1, 2, 3, and 4.
Among possible candidates, let us
consider the following three types of color bases.

\begin{enumerate}
\item Primitive base
\begin{equation}
 \left\{
 \begin{array}{ll}
   C_1 &= \delta_{12}\delta_{34} \\
   C_2 &= \delta_{14}\delta_{32} 
 \end{array}\right.
\end{equation}
\item Orthogonal base
\begin{equation}
 \left\{
 \begin{array}{ll}
   C_1^{(o)} &= C_1+C_2 \\
   C_2^{(o)} &= C_1-C_2 
 \end{array}\right.
\end{equation}
\item Extended orthogonal base
\begin{equation}
 \left\{
 \begin{array}{ll}
   C_1^{(e)} &= C_1 -\delta_{1234} \\
   C_2^{(e)} &= C_2 -\delta_{1234} \\
   C_3^{(e)} &= \delta_{1234}
 \end{array}\right.
\end{equation}
\end{enumerate}
Here  $\delta_{1234}=1$ only when all four indices are equal
and $=0$ otherwise.

For each primitive base, one can assign a configuration
of two strings naturally. They are, however, not diagonal;
the interpretation of interference term is obscure.
In the second case, the orthogonal bases, 
there is no interference term. However, each base
\(C_k^{(o)}\) do not relate directly to the string picture
and each \(C_k^{(o)}\) contains \(|C_1|^2\) and \(|C_2|^2\)
with equal weight.
The same thing happens in elementary
quantum mechanics: The information on
left/right circular polarization is lost
after the light goes through a linear
polarization filter.
In the third example, the first two \(C_k^{(e)}\)
can be assigned to two strings of different colors.
The third one is a kind of 4-quark loop worm.
They are orthogonal to each other but they do not
fit well with the {\tt JETSET} approach.

The situation does not change for the case with more external
particles, and for both cases with or without the gluon
conversion.
We have to say that
there is some {\it distance} between the color bases and
the definition of color singlet sub-system (e.g., strings)
in the hadronization model.

In the present scheme of {\tt grc4f}, we use
the primitive base case and the color flow is
chosen for each event stochastically with    
weight \(C_k C_k^{\dagger} |F_k|^2\).
 
\medskip
\leftline{\large\bf 2.3.3 Options}
\smallskip

There is an option to include the overall
QCD correction factor.  Here the
cross section is multiplied by
a simple factor $(1+\alpha_S/\pi)$ for each quark vertex.

When the final state is a four-quark state,
gluon exchange diagrams are included.   
It is possible to remove these by setting control data.
%
%
%

\medskip
\leftline{\large\bf 2.4 Kinematics and cuts}
\smallskip
%

In the phase-space integration by the adaptive Monte Carlo method, a
proper treatment of the following singuler behavior of the amplitude
is necessary;
\begin{enumerate}
\item $s$-channel $W$, $Z$ resonances and $\gamma$ propagators.
\item $t$-channel electron and electron neutrino propagators
      in $Z$-pair ($Z\gamma$, $\gamma\gamma$) production and
      in $W$-pair production.
\item The electron (positron) forward dcattering  
of the two-photon processes.
\item Mixed resonance of $W$-pair and $Z$-pair 
      ($Z\gamma$, $\gamma\gamma$) processes, such as $u {\bar d}
      {\bar u} d$ process.
\item Identical particles in the final state, such as 
      $\mu^+ \mu^- \mu^+ \mu^-$. 
\end{enumerate}

{\tt grc4f} includes the mapping routine from eight integration 
variables to the four-momenta of final particles with
proper treatment of these singuler behavior. 
The forward scattering of the two-photon processes are not fully treated.
Higgs boson diagrams are included in the amplitude, however the 
kinematics
does not treat the Higgs boson resonance yet.
(For cross
section with very forward electrons or with Higgs boson resonance, 
contact the authors
\footnote{E-mail address:{\tt grc4f@minami.kek.jp}}.) 
For the identical particles 
in the final state, some momentum ordering is assumed in the kinematics
routine, then, for instance, the first particle has always larger
momentum than the second particle.

One can apply some experimental cuts for the phase space integration;
energy and angle cuts on all final particles and a invariant-mass
cut on any pair of final particles.
These parameters can be modified by the user. For comparison
with other programs, {\tt grc4f} has the 
{\tt canonical cut} option used in ref.\cite{yellowreport}, which is;

\begin{enumerate}
\item the energy of charged leptons must be greater than 1 GeV.
\item the polar angle of charged leptons must be between
      10 and 170 degree.
\item the energy of quarks must be greater than 3 GeV.
\item the opening angle between charged leptons and quarks must be
      greater than 5 degree.
\item the invariant mass of quark pairs must be greater than 5 GeV.
\end{enumerate}

Here the `charged leptons' include $\tau$'s too. 
\par
\medskip
\leftline{\large\bf 2.5 Coulomb correction }
\smallskip
%
This effect has been
originally discussed for the {\sl off-shell} $W$-pair 
production\cite{coulomb}
corresponding to the three resonant diagrams. The
set of these 
diagrams cannot satisfy the gauge invariance and has no physical
meaning in principle. 
In {\tt grc4f}, however, 
all relevant diagrams are taken into account
and the gauge invariance is restored from the contribution of the
non-resonant diagrams 
(besides an effect from the finite width of 
$W$ boson). To maintain the
invariance we adopt the following prescription: {\sl the
Coulomb factor is multiplied to the minimum set of gauge invariant diagrams 
containing the $W$-pair production}\cite{yellowreport}.
Hence the Coulomb multiplicative factor appears in some
diagrams even without $W$-pair. The following formula is used:
\begin{eqnarray*}
\sigma_{Coul}=\sigma_{gauge~inv.}{\alpha \pi \over 2 \bar{\beta}}
  \biggl[ 1-{2\over\pi}{\rm arctan}\biggl(
            {|\beta_M+\Delta|^2-\bar{\beta}\over
             2\bar{\beta}{\rm Im}\beta_M } \biggr) \biggl],
\end{eqnarray*} 
where
\begin{eqnarray*} 
\bar{\beta}&=&{1\over s}\sqrt{s^2-2s(k^2_++k^2_-)+(k^2_+-k^2_-)^2},
    \nonumber \\ 
\beta_M&=&\sqrt{1-4M^2/s},~~~~M^2=M^2_W-iM_W\Gamma_W,
~~~~~ \Delta={|k^2_+-k^2_-| \over s},
\end{eqnarray*} 
and $-\pi/2<{\rm arctan}y<\pi/2$. Here $\bar{\beta}$ is the average
velocity of the $W$ boson in {\sl its center-of-mass system}.
Two squared momentum $k^2_+$ and $k^2_-$ are the virtualities of 
the intermediate $W$ bosons.
\par
\medskip
\leftline{\large\bf 2.6 Anomalous coupling }
\smallskip
%
In the program, the anomalous coupling of heavy boson is available for 
the convenience of the user who may
be interested to see the result of an {\sl ad hoc} standard model extension,
although there is no 
definite and reliable model beyond the standard model at present.
The program includes only those terms which conserve C and P 
invariance which correspond to the following effective 
Lagrangian\cite{hagiwara};
\begin{eqnarray*}
L_{eff}&=&-ig_V(W^{\dag}_{\mu \nu} W^{\mu} V^{\nu}-
                  W^{\dag}_{\mu} V_{\nu} W^{\mu \nu})
-ig_V\kappa_VW^{\dag}_{\mu}W_{\nu}V^{\mu \nu}
-ig_V \frac{\lambda_V}{m_W^2}W^{\dag}_{\lambda \mu} 
           W^{\mu}_{\nu}V^{\lambda\nu}, \\
 W_{\mu\nu}&=&\partial_{\mu}W_{\nu}-\partial_{\nu}W_{\mu}, ~~~ 
 V_{\mu\nu}=\partial_{\mu}V_{\nu}-\partial_{\nu}V_{\mu}, ~~~
 V=Z^0 \hbox{ or } \gamma, ~~~
g_V = \left\{ \begin{array}{ll}
         -e & V = \gamma \\
         -e {\rm cot}\theta_W & V = Z^0 
              \end{array} \right.
\end{eqnarray*}
Here
$\kappa_V$ and $\lambda_V$ stand for the anomalous couplings, which 
takes 1 and 0, respectively, in the standard model. Deviation from
these values corresponds to the introduction of 
anomalous coupling with arbitrary strength.

%% file: chap2.tex
\medskip
\leftline{\large\bf  3. Structure of the program}
\smallskip

The generator {\tt grc4f} enables us to generate any 
$e^+e^- \rightarrow$ 4-fermions events with or without radiative corrections.
This requires different calculation in some part of the program.
In addition there are many options covering theoretical and experimental
requirements.
Since all program components are  distributed as source code, 
users can set all options by editing the relevant subprograms directly.
However, 
an interface program {\tt grc4f} has been created to lighten
the user's burden. It 
selects and/or corrects the components which are affected by the various 
options and create a "Makefile" according to the user requirements.
This procedure is called the source generation phase.

In the integration step, the matrix element of a selected
process is integrated over the phase space by the subprogram {\tt BASES},
which gives the effective total and differential cross sections and
the probability distribution
used in event generation phase.
There, the subprogram {\tt SPRING} samples a point
in the phase space and test if it can be accepted as a new event
according to its probability.
When an event is accepted the program control returns from {\tt SPRING}
to the main program, where further analysis is performed by using
the resultant four-momenta of the final state particles.

There are therefore, three steps in the generator {\tt grc4f}, the first is
the source generation, the second is the integration, and the third is
the event generation.
\par
\noindent
In addition to the user interface program {\tt grc4f}, the following
programs are available in the {\tt grc4f} package:
\begin{itemize}
\item[i)] The main programs {\tt MAINBS} and {\tt MAINSP}, and
all program components for the integration and event generation steps.
\item[ii)] The interface programs to CERNLIB ({\tt GRC2CL}) and to
{\tt JETSET} ({\tt GRC2SH}).
\item[iii)] 76 function programs {\tt FUNC}s for processes
$e^+e^-$ $\rightarrow$ 4-fermions, each of which calculates the numerical
value of the differential cross section for each process.
\item[iv)] The kinematics subprograms {\tt KINMOQ} and {\tt KINEMO} for
the processes with radiative correction using the
QED parton shower model or the electron structure function,
respectively.
\item[v)] The library {\tt CHANEL}\cite{chanel} used to calculate
the numerical values of basic components of Feynman 
diagrams in terms of helicity amplitudes.
\item[vi)] The numerical integration and event generation program package
{\tt BASES/SPRING v5.1}\cite{bases}.
\end{itemize}
\par\noindent
The relationship among these program components and their function 
are presented in the next two sections.
\par
\medskip
\leftline{\large\bf 3.1 Source generation step}
\smallskip
The user interface program {\tt grc4f} reads the parameters from
control data prepared by users, which contains
process selection, type of radiative corrections, physical options, experimental cuts,
etc., and generates all necessary components:
\par
\begin{itemize}
\item[i)] A main program {\tt MAINBS} for the integration.
\item[ii)] Interface programs {\tt GRC2CL} and {\tt GRC2SH}.
\item[iii)] Three initialization subprograms {\tt USRPRM}, {\tt MODMAS}
          and {\tt KINIT}, and
\item[iv)] a ``Makefile''.
\end{itemize}

\medskip
\leftline{\large\bf 3.2 Integration step}
\smallskip
\par
Before starting the numerical integration the main program {\tt MAINBS}
invokes an initialization subprogram {\tt USERIN}, where the following
subprograms are called in this order:
\par
\smallskip
\begin{tabular}{llp{30em}}
{\tt USRPRM} & : & To define the set of so-called 
                 "canonical cut"~\cite{yellowreport}
                 authorized by LEP200 working
                 group and some additional optional parameters. \\
{\tt SETMAS} & : & To set masses and decay widths of particles. \\
{\tt MODMAS} & : & To alter the default values of all parameters defined
in {\tt SETMAS}.\\
{\tt AMPARM} & : & To set the coupling constants and others parameters.\\
{\tt KINIT}  & : & To set the parameters for the integration, kinematics,
cuts, histograms etc.
\end{tabular}
\par
The subprograms {\tt SETMAS} and {\tt AMPARM} are generated by the {\tt GRACE}
system.
There are no consistency checks among the constants, e.g.~$M_Z$, $M_W$
and $\sin\theta_W$, so all modifications on these parameters in
the subprograms {\tt USRPRM}, {\tt MODMAS} and {\tt KINIT} are under the
user responsibility alone.
\par
The integration program {\tt BASES} calculates the scattering
cross section by sampling the function {\tt FUNC} on the allowed
phase space segmented by an self adapted grid where finer
cells are clustered on the high gradient zones. This is an iterative
process running until either the maximum number of allowed iteration is reached
or the required accuracy is obtained.
In the function {\tt FUNC}, the kinematics subroutine {\tt KINEMO} or {\tt KINMOQ} 
is used to map the integral variables with the four-momenta of the final state 
particles.
{\tt KINEMO} is used for reaction with no radiative corrections or those 
involving initial radiation treated with the structure function
techniques. {\tt KINMOQ} is called for processes where radiative corrections
are computed with the QED parton shower method.
The subprograms {\tt AMPTBL} and {\tt AMPSUM} are further called 
for calculating the helicity amplitudes and the sum of the matrix 
element squared.

{\bf It is recommended to look at the integration result carefully, 
especially over the convergency behaviors both 
for the grid optimization and integration steps.}
When the accuracy of each iteration fluctuates, iteration by iteration, and,
in some case, it may jump up suddenly to a large value compared to the other
iterations, the resultant estimate of integral may not be reliable.
There are two possible origins of this behavior; too few
sampling points or/and an unsuitable choice of the kinematics.

After the numerical integration by {\tt BASES}, the subprograms {\tt BSINFO} 
and {\tt BHPLOT} are called to print the result of integration and
the histograms, respectively.
Before terminating the integration procedure the probability distribution
obtained by the integration can be saved in a file by invoking {\tt BSWRIT}, 
which is then used for the event generation by {\tt SPRING}.
\par
\medskip
\leftline{\large\bf 3.3 Event generation step}
\smallskip
\par
After integrating the differential cross section and saving the probability
distribution, the main program, {\tt MAINSP}, handles the event generation 
program.
The subprogram {\tt BSREAD} is invoked to restore
the probability distribution and then the subprogram {\tt USERIN} is called.
Each call to {\tt SPRING} generates one event by sampling a point in the phase
volume. It then
calculates the differential cross section at that point using 
the same function {\tt FUNC} seen in the 
integration phase.
Using the usual unweighting technique, this event is accepted or rejected.
When an event is generated, {\tt SPRING} returns 
particle identification and four momenta.
The event information is stored in the labelled common {\tt LUJETS} by
calling subprogram {\tt SP2LND}.
Here, the information for the color connection to be
referred by {\tt JETSET} is also supplied.
Among the final states in Appendix B, those in Table 3
have non-trivial color flow
which is determined as is described in the section 2.3.

Then parton shower and hadronization of quarks and gluons can be performed
by calling {\tt LUSHOW} and {\tt LUEXEC}.
At the end of the event generation, the routine {\tt SPINFO} and {\tt SHPLOT}
are invoked successively for printing event generation and
histograms.


%% file: how.tex
\leftline{\large\bf 4. How to run the program}

The user should first prepare control data to define the process, 
options and parameters for experimental cuts etc.
The user interface program {\tt grc4f} takes this control data 
(let's call it {\tt control.data}) as an input.
\par\smallskip
{\tt
\begin{tabular}{lll}
process & = & eNEuD \\
energy  & = & 190.0d0 \\
canon   & = & yes \\
type    & = & tree \\
massive & = & yes \\
coulomb & = & no \\
anomal  & = & no \\
qcdcr   & = & no \\
end & &  
\end{tabular}
}
\par\smallskip\noindent
The first line specifies the process to be calculated and
the second is the center of mass energy in GeV unit.
The others are options, whose meanings are given in Appendix A.
Then the user may type:
\par
\begin{verbatim}
% grc4f < control.data 
\end{verbatim}
\par\noindent
If the message ``syntax error'' is returned,
no files will be generated and the contents of the control data
file must carefully be checked.
After a successful completion, the following messages should be returned:
\par
\begin{verbatim}
Process is "eNEuD"
Energy is "190.0d0"
CANON <yes>
MASSIV <yes>
COULMB <no>
ANOMAL <no>
QCDCR <no>
bye-bye
absolute directory name is /home/grc4f/prc/elNEuqDQ
------------------------------------------
cd /home/grc4f/prc/elNEuqDQ
make
integ
spring
------------------------------------------
\end{verbatim}
\par\noindent
According to the parameters given in control data, 
the files, i.e.~ {\it usrprm.f}, 
{\it modmas.f}, {\it kinit.f},
{\it mainbs.f}, {\it mainsp.f} and {\it Makefile}, are generated in
a specified subdirectory ( {\tt elNEuqDQ} in this case).
\par\noindent
According to the last four lines in the message, 
users can proceed the calculations as follows:
\par
\begin{itemize}
\item[i)] Change directory by
\begin{verbatim}
% cd /home/grc4f/prc/elNEuqUD
\end{verbatim}
\item[ii)] Create an executable {\tt integ} for the integration by typing:
\begin{verbatim}
% make integ
\end{verbatim}
\par
\item[iii)] Numerical integration is actually performed by typing:
\begin{verbatim}
%  integ
\end{verbatim}
\par
The results of
integration step are displayed as well as written in an
output file \\ 
{\tt bases.result}. The total cross section in $pb$ with
the error are displayed at the last row, under {\tt Cumulative Result}, 
in the table of the {\tt Convergence Behavior for the Integration step}.
The differential cross sections as a function of the energy and
angle of each particle and invariant masses of any two final particles
will also be printed. The probability distribution is written in a file 
{\tt bases.data} which will be used in the event generation step by {\tt spring}.

\item[iv)] Before running the event generation, users should edit 
{\tt mainsp.f} to set additional parameters like the number of events 
and call user's own analysis routines.

The following is the structure of the generated {\tt mainsp.f}, where
four-momenta of all particles are stored in the common/lujets/ in the 
{\tt JETSET} format by calling subprogram {\tt sp2lnd} in the event-loop:
\begin{verbatim}
      implicit real*8(a-h,o-z)
       ....................
      real*4  p,v
      common/lujets/n,k(4000,5),p(4000,5),v(4000,5)
       ....................
      mxtry  = 50
      mxevnt = 10000
       ....................
      do 100 nevnt = 1, mxevnt

         call spring( func, mxtry )
       ....................
*         -----------------
           call sp2lnd
*         -----------------
*
*       =================================
*       (   user_analysis   )
*       =================================
*
  100 continue
       ....................
      stop
      end
\end{verbatim}

\item[v)] Create an executable {\tt spring} for event generation by typing:

\begin{verbatim}
% make spring
\end{verbatim}

\item[vi)] Start the event generation by typing:

\begin{verbatim}
% spring
\end{verbatim}

Information of the event generation will be written in the
{\tt spring.result} file. Users should pay special attention to the 
histograms. The distributions of generated events are superimposed
by ``{\tt 0}'' on the original histograms by {\tt BASES}. These two
distributions should be consistent within the statistical error of
the generation. For the detail of the output files of {\tt BASES}
and {\tt SPRING},
user can consult the Ref.\cite{bases}. 

\end{itemize}



%% file: remark.tex
\leftline{\large\bf 5. Summary}


The generator {\tt grc4f} enables us to calculate the effective cross 
section and to generate events for one of 76 $e^+e^-$ $\rightarrow$ 
4-fermions processes listed in appendix B. 
It is dedicated to the LEP and future linear collider physics studies.
The produced
quarks can be hadronized according to {\tt JETSET}. Also
processes with initial and final radiations can be 
generated in terms of the electron structure function or the QED 
parton shower methods, though the inclusion of the interference between
the initial and final radiations requires further study. 

There still remain important problems to be solved and further
necessary 
improvements. Among them it will be desirable to extend the program
so as to produce several final state processes in a single run like 
$e^+e^- \to$~4-quarks. In order to get more precise QED corrections 
an implementation of a complete next-to-leading logarithmic 
effects\cite{nll} is unavoidable. 

In this version, we have assumed that the hadronization of partons
takes place independent of the hard interaction which produces partons.
However a very important contribution arises from 
diagram where a gluon is exchanged between quarks produced in the
two $W$ decays. 
Taking into account this effect implies
higher order calculations or the implementation of a specific
phenomenologic correction\cite{yellowreport}.

Finally the present version provides only 76 processes, 
the missing processes will be prepared 
soon by using {\tt GRACE}. 


%% file: ack.tex
\leftline{\large\bf Acknowledgements}

The authors would like to thank G. Coignet, F. Boudjema, B. Mele
for their interest and encouragement and colleagues in Minami-Tateya
group of KEK for their help. This work was done in the KEK-LAPP
collaboration supported in part by Mombusho in Japan 
under the Grant-in-Aid for International Scientific Research Program 
No.07044097,
and CNRS/IN2P3
in France.

%% file: appa.tex
\leftline{\large\bf Appendix A. Option table in Control data }

In the table below, the default values are underlined 
and  the relation between commands and variable/array in
Fortran sources is also described. Variable names are written in
bold letters and file names in italic. 

\begin{itemize}
\item[i)] Process selection.

\begin{tabular}{|p{4.2em}cp{25em}|} \hline
{\tt Process}&{\tt =}& \underline{\tt eNEuD} \\
 & & abbreviation of process name \\ \hline
\end{tabular}

This specifies the subdirectory name, where the necessary function program 
to calculate matrix elements are stored.
Tables 1, 2, and 3 in Appendix B shows the abbreviation
of process names and the subdirectory names where they are stored are listed.

\item[ii)] Center of mass energy.

\begin{tabular}{|p{4.2em}cp{25em}|} \hline
{\tt energy}&{\tt =}& \underline{\tt 190.d0} \\
 & & CMS energy in GeV \\ 
& &{\tt w} in {\it kinit.f} \\ \hline
\end{tabular}

\item[iii)] Global options

\begin{tabular}{|p{4.2em}cp{25em}|} \hline
{\tt helicity1}   &{\tt =}& {\tt \underline{average}},
~{\tt left}, ~{\tt right}\\
 & & Helicity state for the initial electron.\\
{\tt helicity2}   &{\tt =}& {\tt \underline{average}},
~{\tt left}, ~{\tt right}\\
 & & Helicity state for the initial positron.\\
{\tt helicity3}   &{\tt =}& {\tt \underline{sum}},
~{\tt left}, ~{\tt right}\\
 & & Helicity state for 3rd particle.\\
{\tt helicity4}   &{\tt =}& {\tt \underline{sum}},
~{\tt left}, ~{\tt right}\\
 & & Helicity state for 4th particle.\\
{\tt helicity5}   &{\tt =}& {\tt \underline{sum}},
~{\tt left}, ~{\tt right}\\
 & & Helicity state for 5th particle.\\
{\tt helicity6}   &{\tt =}& {\tt \underline{sum}},
~{\tt left}, ~{\tt right}\\
 & & Helicity state for 6th particle.\\ \hline
\end{tabular}

\begin{tabular}{|p{4.2em}cp{25em}|} \hline
{\tt type}   &{\tt =}& {\tt \underline{tree}}, ~{\tt sf}, 
~{\tt qedpsi}, ~{\tt qedpsif}\\
& & Type of calculation: \\
 & & Without radiation({\tt tree}), 
ISR with structure function({\tt sf}),
ISR with QEDPS({\tt qedpsi}) and 
ISR and FSR with QEDPS({\tt qedpsif}). \\
& & {\tt jqedps = 0}, {\tt isr = 0} without radiation({\tt tree}).\\
& & {\tt jqedps = 0}, {\tt isr = 1} for {\tt sf}.\\
& & {\tt jqedps = 1}, {\tt ips = 1} for {\tt qedpsi}.\\
& & {\tt jqedps = 1}, {\tt ips = 2} for {\tt qedpsif}.\\
& & {\tt jqedps} in {\it usrprm.f}, {\tt isr} 
and {\tt ips} in {\it kinit.f}. \\ \hline
\end{tabular}

\begin{tabular}{|p{4.2em}cp{25em}|} \hline
{\tt canon}   &{\tt =}& {\tt \underline{yes}}, ~{\tt no} \\
& & Apply canonical cuts or not:\\
& & {\tt jcanon = \underline{1}} or {\tt 0} in {\it usrprm.f}.\\ \hline
{\tt massiv}   &{\tt =}& {\tt \underline{yes}}, ~{\tt no} \\
& & Quarks are massive or massless:\\
& & {\tt jqmass = \underline{1} or 0} in {\it usrprm.f}.\\ \hline
{\tt width }   &{\tt =}& {\tt \underline{run}}, ~{\tt fixed} \\
& & Running width or fixed width:\\
& & {\tt jwidth = \underline{0}} or {\tt 1} in {\it usrprm.f}.\\ \hline
\end{tabular}

\item[iv)] Physical options

\begin{tabular}{|p{4.2em}cp{25em}|} \hline
{\tt coulomb}   &{\tt =}& {\tt yes}, ~{\tt \underline{no}} \\
& & Coulomb correction: \\
& & {\tt jcolmb = 1} or {\tt \underline{0}} in {\it modmas.f}.\\ \hline
{\tt qcdcr}   &{\tt =}& {\tt yes}, ~\underline{no} \\
& & Include overall QCD correction factor: \\
& & {\tt jqcdcr = 1} or {\tt \underline{0}} in {\it modmas.f}.\\ \hline
{\tt gluon}   &{\tt =}& {\tt yes}, ~{\tt \underline{no}} \\
& & Include diagrams with gluon exchange:\\
& & {\tt jgluon = 1} or {\tt \underline{0}} in {\it modmas.f}.\\ \hline
\end{tabular}

\begin{tabular}{|p{4.2em}cp{25em}|} \hline
{\tt anomal}   &{\tt =}& {\tt yes}, ~{\tt \underline{no}} \\
& & {\tt jano3v = 1} or {\tt \underline{0}} in {\it modmas.f}.\\ 
& & Anomalous coupling: \\ \hline
{\tt ankaa}   &{\tt =}& {\tt \underline{1.0d0}}\\
 & & $\kappa_{\gamma}$.\\
{\tt anlma}   &{\tt =}& {\tt \underline{0.0d0}}\\
 & & $\lambda_{\gamma}$.\\
{\tt ankaz}   &{\tt =}& {\tt \underline{1.0d0}}\\
 & & $\kappa_{Z^{0}}$.\\
{\tt anlmz}   &{\tt =}& {\tt \underline{0.0d0}}\\
 & & $\lambda_{Z^{0}}$. \\
& &See section 2.6 for the definition of
above variables.\\ 
& &if ``{\tt anomalous = no}'' is specified, these options
affect nothing. \\ \hline
\end{tabular}

\item[v)] Masses, widths and $\alpha$ (only if ``{\tt canon = no}'').

\begin{tabular}{|p{4.2em}cp{25em}|} \hline
{\tt amw}   &{\tt =}& {\tt \underline{80.23D0}} \\
 & & $W$ mass in GeV: {\tt amw} in {\it modmas.f}.\\ \hline
{\tt agw}   &{\tt =}& {\tt \underline{2.03367033062746D0}} \\
 & & $W$ width in GeV:{\tt agw} in {\it modmas.f}.\\ \hline
{\tt amz}   &{\tt =}& {\tt \underline{91.1888D0}} \\
 & & $Z$ mass in GeV: {\tt amz} in {\it modmas.f}.\\ \hline
{\tt agz}   &{\tt =}& {\tt \underline{2.4974D0}} \\
 & & $Z$ width in GeV: {\tt agz} in {\it modmas.f}.\\ \hline
{\tt alphai}   &=& {\tt \underline{128.07D0}} \\
 & & $\alpha^{-1}$:{\tt alphai} in {\it modmas.f}.\\ \hline
{\tt alpha\_s}   &=& {\tt \underline{0.12D0}} \\
 & & $\alpha_s$:{\tt alpha\_s} in {\it modmas.f}.\\ \hline
\end{tabular}


\item[vi)] Experimental cuts (only if ``{\tt canon = no}'').

The numbering convention of particles  follows the {\tt GRACE} scheme, where
the initial electron and positron are 1st and 2nd, respectively, and
the four final particles are numbered 3, 4, 5, 6.
In the process name of Table 1, 2, and 3,
the order of particles corresponds to this numbering convention.
For instance, in the process, $e^- \bar{\nu_{e}} u \bar{d}$,
the 3rd is $e^-$, the 4th is $\bar{\nu_{e}}$,
the 5th is $u$ and the 6th is $\bar{d}$.

\begin{tabular}{|p{4.2em}cp{25em}|} \hline
{\tt thecut3}   &{\tt =}& {\tt \underline{180.d0},\underline{0.0d0}} \\
 & & Angle cut for 3rd particle in degree 
(backward-angle,forward-angle).\\
 & & {\tt coscut(1:2,1)=cos(thecut3)} in {\it kinit.f}.\\ \hline
{\tt thecut4}   &{\tt =}& {\tt \underline{180.d0},\underline{0.0d0}} \\
 & & Angle cut for 4th particle in degree 
(backward-angle,forward-angle).\\
 & & {\tt coscut(1:2,2)=cos(thecut4)} in {\it kinit.f}.\\ \hline
{\tt thecut5}   &{\tt =}& {\tt \underline{180.d0},\underline{0.0d0}} \\
 & & Angle cut for 5th particle in degree
(backward-angle,forward-angle).\\
 & & {\tt coscut(1:2,3)=cos(thecut5)} in {\it kinit.f}.\\ \hline
{\tt thecut6}   &{\tt =}& {\tt \underline{180.d0},\underline{0.0d0}} \\
 & & Angle cut for 6th particle in degree 
(backward-angle,forward-angle).\\
 & & {\tt coscut(1:2,4)=cos(thecut6)} in {\it kinit.f}.\\ \hline
\end{tabular}

Instead of giving a numerical value, the user can use the keywords as 
below:
{\tt amass1(}{\it i}{\tt )} has the mass for {\it i}-th particle and
{\tt w} is the CM energy.

\begin{tabular}{|p{4.2em}cp{25em}|} \hline
{\tt engcut3}   &{\tt =}& {\tt \underline{amass1(3)},\underline{w}} \\
 & & Energy cut for 3rd particle (min.,max.) \\
 & & {\tt engyct(1:2,1)} in {\it kinit.f} \\ \hline
{\tt engcut4}   &{\tt =}& {\tt \underline{amass1(4)},\underline{w}} \\
 & & Energy cut for 4th particle (min.,max.) \\
 & & {\tt engyct(1:2,2)} in {\it kinit.f} \\ \hline
{\tt engcut5}   &{\tt =}& {\tt \underline{amass1(5)},\underline{w}} \\
 & & Energy cut for 5th particle (min.,max.) \\
 & & {\tt engyct(1:2,3)} in {\it kinit.f} \\ \hline
{\tt engcut6}   &{\tt =}& {\tt \underline{amass1(6)},\underline{w}} \\
 & & Energy cut for 6th particle (min.,max.) \\
 & & {\tt engyct(1:2,4)} in {\it kinit.f} \\ \hline
\end{tabular}

\begin{tabular}{|p{4.2em}cp{25em}|} \hline
{\tt ivmcut34}   &{\tt =}& {\tt \underline{amass1(3)+amass1(4)}},
{\tt \underline{w-(amass1(5)+amass1(6))}}\\
& & Invariant mass cut for 3-4 particles(min.,max.) \\
& & {\tt amasct(1:2,1)} in {\it kinit.f} \\ \hline
{\tt ivmcut56}   &{\tt =}& {\tt \underline{amass1(5)+amass1(6)}},
{\tt \underline{w-(amass1(3)+amass1(4))}} \\
& & Invariant mass cut for 5-6 particles(min.,max.) \\
 & & {\tt amasct(1:2,2)} in {\it kinit.f} \\ \hline
{\tt ivmcut35}   &{\tt =}& {\tt \underline{amass1(3)+amass1(5)}},
{\tt \underline{w-(amass1(4)+amass1(6))}} \\
& & Invariant mass cut for 3-5 particles(min.,max.) \\
 & & {\tt amasct(1:2,3)} in {\it kinit.f} \\ \hline
\end{tabular}

\begin{tabular}{|p{4.2em}cp{25em}|} \hline
{\tt ivmcut46}   &{\tt =}& {\tt \underline{amass1(4)+amass1(6)}},
{\tt \underline{w-(amass1(3)+amass1(5))}} \\
& & Invariant mass cut for 4-6 particles(min.,max.) \\
 & & {\tt amasct(1:2,4)} in {\it kinit.f} \\ \hline
{\tt ivmcut36}   &{\tt =}& {\tt \underline{amass1(3)+amass1(6)}},
{\tt \underline{w-(amass1(4)+amass1(5))}} \\
& & Invariant mass cut for 3-6 particles(min.,max.) \\
 & & {\tt amasct(1:2,5)} in {\it kinit.f} \\ \hline
{\tt ivmcut45}   &{\tt =}& {\tt \underline{amass1(4)+amass1(5)}},
{\tt \underline{w-(amass1(3)+amass1(6))}} \\
 & & Invariant mass cut for 4-5 particles(min.,max.) \\
 & & {\tt amasct(1:2,6)} in {\it kinit.f} \\ \hline
\end{tabular}


\item[vii)] Parameters for integration.

\begin{tabular}{|p{4.2em}cp{25em}|} \hline
{\tt itmx}   &{\tt =}& {\tt \underline{7}},~{\tt \underline{15}} \\
 & & Iteration numbers: {\tt itmx1}, {\tt itmx2} in {\it kinit.f} \\
{\tt acc}   &=& {\tt \underline{0.1}}, ~{\tt \underline{0.05}} \\
 & & Accuracies in \%:{\tt acc1}, {\tt acc2}
 in {\it kinit.f} \\
{\tt ncall}   &=& {\tt \underline{40000}} \\
 & & Sampling points: {\tt ncall}
 in {\it kinit.f} \\ \hline
\end{tabular}

\item[viii)] Parameters for event generation.

\begin{tabular}{|p{4.2em}cp{25em}|} \hline
{\tt mxtry}   &=& \underline{50} \\
& & Maximum trial numbers: {\tt mxtry}
 in {\it mainsp.f}\\
{\tt mxevnt}   &=& \underline{10000} \\
& & Maximum event numbers: {\tt mxevnt}
 in {\it mainsp.f}\\
{\tt hadron}   &=& yes, \underline{no} \\
& & Hadronization with {\tt JETSET}\cite{jetset} for event generation:\\
& & If {\tt yes}, then a statement, {\tt call luexec} 
is added in {\it mainsp.f}. In this case, program {\tt JETSET}
is necessary, and one should specify {\tt JETSET}
file location in this control data. \\ \hline
{\tt jetset}  &{\tt =}& {\tt /home/jetset/jetset74.o} \\
& & Object file name for {\tt JETSET},
if {\tt hadron = yes} \\ \hline
{\tt qcd\_shower}  &=& yes, \underline{no} \\
& & QCD parton shower with {\tt JETSET}\cite{jetset} for event 
generation:\\
& & If {\tt yes}, then a statement, {\tt call lushow}
is added in {\it mainsp.f}. In this case, program {\tt JETSET} is
necessary, and one should specify {\tt JETSET} file location in this
control data. \\ \hline
\end{tabular}

Those parameters affect {\it Makefile}.

\item[ix)] HBOOK interface.

\begin{tabular}{|p{4.2em}cp{25em}|} \hline
{\tt cernlib}   &=& yes, \underline{no} \\
& & If {\tt yes}, then the {\it bases.hbook} in integration 
and \par
{\it spring.hbook} in event generation will be generated. \\ \hline
\end{tabular}

Those parameters affect {\it mainbs.f}, {\it mainsp.f} and
{\it Makefile}.

\item[x)] End of description.

\begin{tabular}{|p{4.2em}cp{25em}|} \hline
{\tt end}   && \\ \hline
\end{tabular}

After the command {\tt end} any command is neglected.
\end{itemize}

%% file: tbl.tex
\newcommand{\nue}{{\nu_e}}
\newcommand{\nuebar}{{\bar{\nu}_e}}
\newcommand{\numu}{{\nu_\mu}}
\newcommand{\numubar}{{\bar{\nu}_\mu}}
\newcommand{\nutau}{{\nu_\tau}}
\newcommand{\nutaubar}{{\bar{\nu}_\tau}}
\newcommand{\electron}{{e^-}}
\newcommand{\positron}{{e^+}}
\newcommand{\muon}{{\mu^-}}
\newcommand{\antimuon}{{\mu^+}}
\newcommand{\taum}{{\tau^-}}
\newcommand{\antitau}{{\tau^+}}
\newcommand{\uq}{{u}}
\newcommand{\ubar}{{\bar{u}}}
\newcommand{\dq}{{d}}
\newcommand{\dbar}{{\bar{d}}}
\newcommand{\cq}{{c}}
\newcommand{\cbar}{{\bar{c}}}
\newcommand{\sq}{{s}}
\newcommand{\sbar}{{\bar{s}}}
\newcommand{\bq}{{b}}
\newcommand{\bbar}{{\bar{b}}}

\leftline{\large\bf Appendix B. Process table}

\vspace{2mm}

Here, the 76 processes included in {\tt grc4f} are listed.
In the heading, 'abbrev.' and 'dir.' stand for the
abbreviated name used in the control card and directory 
name where the generated code is stored, respectively.

\vspace{2mm}

\begin{center}
\begin{tabular}{|l|l|l||l|l|l|} \hline
process & abbrev. & dir. &
process & abbrev. & dir. \\ \hline \hline
\rule{0mm}{4mm}
$\electron \nuebar \positron \nue$ & {\tt eNEEne} & {\tt elELneNE} & $\electron \nuebar \numu \antimuon$& {\tt eNEnmMU} & {\tt elNEMUnm} \\ \hline 
\rule{0mm}{4mm}
$\electron \nuebar \nutau \antitau$ & {\tt eNEntTAU} & {\tt elNETAnt} & $\muon \antimuon \numu \numubar $ & {\tt muMUnmNM} & {\tt muMUnmNM} \\ \hline 
\rule{0mm}{4mm}
$\taum \antitau \nutau \nutaubar$ & {\tt tauTAUntNt} & {\tt taTAntNT} & $\muon \numubar \antitau \nutau $ & {\tt muNMTAUnt} & {\tt muNMTAta} \\ \hline 
\rule{0mm}{4mm}
$\electron \positron \electron \positron$ & {\tt eEeE} & {\tt elELelEL} & $\electron \positron \muon \antimuon $ & {\tt eEmuMU} & {\tt elELmuMU} \\ \hline 
\rule{0mm}{4mm}
$\electron \positron \taum \antitau$ & {\tt eEtauTAU} & {\tt elELtaTA} & $\muon \antimuon \muon \antimuon $ & {\tt muMUmuMU} & {\tt muMUmuMU} \\ \hline 
\rule{0mm}{4mm}
$\taum \antitau \taum \antitau$ & {\tt tauTAUtauTAU} & {\tt taTAtaTA} & $\muon \antimuon \taum \antitau $ & {\tt muMUtauTAU} & {\tt muMUtaTA} \\ \hline 
\rule{0mm}{4mm}
$\electron \positron \numu \numubar$ & {\tt eEnmNM} & {\tt elELnmNM} & $\electron \positron \nutau \nutaubar $ & {\tt eEntNT} & {\tt elELntNT} \\ \hline 
\rule{0mm}{4mm}
$\nue \nuebar \muon \antimuon$ & {\tt neNEmuMU} & {\tt neNEmuMU} & $\nue \nuebar \taum \antitau $ & {\tt neNEtauTAU} & {\tt neNEtaTA} \\ \hline 
\rule{0mm}{4mm}
$\nutau \nutaubar \muon \antimuon$ & {\tt ntNTmuMU} & {\tt ntNTmuMU} & $\numu \numubar \taum \antitau $& {\tt nmNMtauTAU} & {\tt nmNMtaTA} \\ \hline 
\rule{0mm}{4mm}
$\nue \nuebar \nue \nuebar$ & {\tt neNEneNE} & {\tt neNEneNE} & $\nue \nuebar \numu \numubar $ & {\tt neNEnmNM} & {\tt neNEnmNM} \\ \hline 
\rule{0mm}{4mm}
$\nue \nuebar \nutau \nutaubar$ & {\tt neNEntNT} & {\tt neNEntNT} & $\numu \numubar \numu \numubar $ & {\tt nmNMnmNM} & {\tt nmNMnmNM} \\ \hline 
\rule{0mm}{4mm}
$\nutau \nutaubar \nutau \nutaubar$ & {\tt ntNTntNT} & {\tt ntNTntNT} & $\nutau \nutaubar \numu \numubar $ & {\tt nmNMntNT} & {\tt ntNTnmNM} \\ \hline 
\end{tabular}
\end{center}

\begin{center}
Table 1 \ Leptonic processes in {\tt grc4f}.
\end{center}

\vspace{3mm}

\begin{center}
\begin{tabular}{|l|l|l||l|l|l|} \hline
process & abbrev. & dir. &
process & abbrev. & dir. \\ \hline \hline
\rule{0mm}{4mm}
$\electron \nuebar \uq \dbar$& {\tt eNEuD} & {\tt elNEuqDQ} & 
$\electron \nuebar \cq \sbar$& {\tt eNEcS} & {\tt elNEcqSQ} \\ \hline 
\rule{0mm}{4mm}
$\muon \numubar \uq \dbar$& {\tt muNMuD} & {\tt muNMuqDQ} & 
$\muon \numubar \cq \sbar$& {\tt muNMcS} & {\tt muNMcqSQ} \\ \hline 
\rule{0mm}{4mm}
$\taum \nutaubar \uq \dbar$& {\tt tauNTuD} & {\tt taNTuqDQ} & 
$\taum \nutaubar \cq \sbar$& {\tt tauNTcS} & {\tt taNTcqSQ} \\ \hline 
\rule{0mm}{4mm}
$\electron \positron \uq \ubar$& {\tt eEuU} & {\tt elELuqUQ} & 
$\electron \positron \cq \cbar$& {\tt eEcC} & {\tt elELcqCQ} \\ \hline 
\rule{0mm}{4mm}
$\electron \positron \dq \dbar$& {\tt eEdD} & {\tt elELdqDQ} & 
$\electron \positron \sq \sbar$& {\tt eEsS} & {\tt elELsqSQ} \\ \hline 
\rule{0mm}{4mm}
$\electron \positron \bq \bbar$& {\tt eEbB} & {\tt elELbqBQ} & 
$\muon \antimuon \uq \ubar$& {\tt muMUuU} & {\tt muMUuqUQ} \\ \hline 
\rule{0mm}{4mm}
$\muon \antimuon \cq \cbar$& {\tt muMUcC} & {\tt muMUcqCQ} & 
$\taum \antitau \uq \ubar$& {\tt tauTAUuU} & {\tt taTAuqUQ} \\ \hline 
\rule{0mm}{4mm}
$\taum \antitau \cq \cbar$& {\tt tauTAUcC} & {\tt taTAcqCQ} & 
$\muon \antimuon \dq \dbar$& {\tt muMUdD} & {\tt muMUdqDQ} \\ \hline 
\rule{0mm}{4mm}
$\muon \antimuon \sq \sbar$& {\tt muMUsS} & {\tt muMUsqSQ} & 
$\muon \antimuon \bq \bbar$& {\tt muMUbB} & {\tt muMUbqBQ} \\ \hline 
\rule{0mm}{4mm}
$\taum \antitau \dq \dbar$& {\tt tauTAUdD} & {\tt taTAdqDQ} & 
$\taum \antitau \sq \sbar$& {\tt tauTAUsS} & {\tt taTAsqSQ} \\ \hline 
\rule{0mm}{4mm}
$\taum \antitau \bq \bbar$& {\tt tauTAUbB} & {\tt taTAbqBQ} & 
$\nue \nuebar \uq \ubar$& {\tt neNEuU} & {\tt neNEuqUQ} \\ \hline 
\rule{0mm}{4mm}
$\nue \nuebar \cq \cbar$& {\tt neNEcC} & {\tt neNEcqCQ} & 
$\nue \nuebar \dq \dbar$& {\tt neNEdD} & {\tt neNEdqDQ} \\ \hline 
\rule{0mm}{4mm}
$\nue \nuebar \sq \sbar$& {\tt neNEsS} & {\tt neNEsqSQ} & 
$\nue \nuebar \bq \bbar$& {\tt neNEbB} & {\tt neNEbqBQ} \\ \hline 
\rule{0mm}{4mm}
$\numu \numubar \uq \ubar$& {\tt nmNMuU} & {\tt nmNMuqUQ} & 
$\numu \numubar \cq \cbar$& {\tt nmNMcC} & {\tt nmNMcqCQ} \\ \hline 
\rule{0mm}{4mm}
$\nutau \nutaubar \uq \ubar$& {\tt ntNTuU} & {\tt ntNTuqUQ} & 
$\nutau \nutaubar \cq \cbar$& {\tt ntNTcC} & {\tt ntNTcqCQ} \\ \hline 
\rule{0mm}{4mm}
$\numu \numubar \dq \dbar$& {\tt nmNMdD} & {\tt nmNMdqDQ} & 
$\numu \numubar \sq \sbar$& {\tt nmNMsS} & {\tt nmNMsqSQ} \\ \hline 
\rule{0mm}{4mm}
$\numu \numubar \bq \bbar$& {\tt nmNMbB} & {\tt nmNMbqBQ} & 
$\nutau \nutaubar \dq \dbar$& {\tt ntNTdD} & {\tt ntNTdqDQ} \\ \hline 
\rule{0mm}{4mm}
$\nutau \nutaubar \sq \sbar$& {\tt ntNTsS} & {\tt ntNTsqSQ} & 
$\nutau \nutaubar \bq \bbar$& {\tt ntNTbB} & {\tt ntNTbqBQ} \\ \hline 
\end{tabular}
\end{center}

\begin{center}
Table 2 \ Semi-hadronic processes in {\tt grc4f}.
\end{center}


\begin{center}
\begin{tabular}{|l|l|l||l|l|l|} \hline
process & abbrev. & dir. &
process & abbrev. & dir. \\ \hline \hline
\rule{0mm}{4mm}
$\uq \dbar \dq \ubar$& {\tt uDUd} & {\tt uqDQdqUQ} & 
$\cq \sbar \sq \cbar$& {\tt cCSs} & {\tt cqSQsqCQ} \\ \hline 
\rule{0mm}{4mm}
$\uq \dbar \sq \cbar$& {\tt uDsC} & {\tt uqDQsqCQ} & 
$\uq \ubar \uq \ubar$& {\tt uUuU} & {\tt uqUQuqUQ} \\ \hline 
\rule{0mm}{4mm}
$\cq \cbar \cq \cbar$& {\tt cCcC} & {\tt cqCQcqCQ} & 
$\dq \dbar \dq \dbar$& {\tt dDdD} & {\tt dqDQdqDQ} \\ \hline 
\rule{0mm}{4mm}
$\sq \sbar \sq \sbar$& {\tt sSsS} & {\tt sqSQsqSQ} & 
$\bq \bbar \bq \bbar$& {\tt bBbB} & {\tt bqBQbqBQ} \\ \hline 
\rule{0mm}{4mm}
$\uq \ubar \cq \cbar$& {\tt uUcC} & {\tt uqUQcqCQ} & 
$\uq \ubar \sq \sbar$& {\tt uUsS} & {\tt uqUQsqSQ} \\ \hline 
\rule{0mm}{4mm}
$\uq \ubar \bq \bbar$& {\tt uUbB} & {\tt uqUQbqBQ} & 
$\cq \cbar \dq \dbar$& {\tt cCdD} & {\tt cqCQdqDQ} \\ \hline 
\rule{0mm}{4mm}
$\cq \cbar \bq \bbar$& {\tt cCbB} & {\tt cqCQbqBQ} & 
$\dq \dbar \sq \sbar$& {\tt dDsS} & {\tt dqDQsqSQ} \\ \hline 
\rule{0mm}{4mm}
$\dq \dbar \bq \bbar$& {\tt dDbB} & {\tt dqDQbqBQ} & 
$\sq \sbar \bq \bbar$& {\tt sSbB} & {\tt sqSQbqBQ} \\ \hline 
\end{tabular}
\end{center}

\begin{center}
Table 3 \ Hadronic processes in {\tt grc4f}.
\end{center}

%% file: install.tex
\leftline{\large\bf Appendix C. Installation}

The source code  is available by {\tt anonymous ftp} from
{\tt ftp.kek.jp} in the directory \\ {\tt kek/minami/grc4f}.
The {\tt grc4f} package contains the complete set of Fortran sources
for 76 processes, the three libraries, i.e., 
{\tt BASES/SPRING}, {\tt CHANEL} and
utilities for kinematics.
Those source codes are written in FORTRAN77.
In addition, {\tt grc4f} provides the
interface program to generate a few Fortran source files
according to the control data specified by the user.
This program is written in C, YACC and LEX.
{\tt grc4f} has been developed on HP-UX, but should run on
any UNIX platform with a fortran complier. 

The procedure of installation is as follows:
\begin{enumerate}
\item Editing {\it Makefile}.

The following macros in {\it Makefile} should be taken care of by 
users themselves.
For example, in the right hand side of {\tt GRC4FDIR} the directory name
where {\tt grc4f} is installed should be given,
and for {\tt FC} and {\tt FOPT} the relevant compiler
name and option for your system should be given.
The other macros can be left as they are.

\begin{tabular}{lcl}
  {\tt GRC4FDIR} &{\tt =}& directory where {\tt grc4f} are installed. \\
  {\tt PRCDIR}   &{\tt =}& directory where process files are installed. \\
 & &  (default is \verb+$(GRC4FDIR)/prc+.) \\
  {\tt LIBDIR}   &{\tt =}& directory where libraries are installed. \\
 & &  (default is \verb+$(GRC4FDIR)/lib+.) \\
  {\tt BINDIR}   &{\tt =}& directory where an executable is installed. \\
 & & (default is \verb+$(GRC4FDIR)/bin+.) \\
  {\tt MACHINE} &{\tt =}& {\tt [hpux|hiux|sgi|dec|sun]} \\
  {\tt FC}      &{\tt =}& FORTRAN compiler command name. \\
  {\tt FOPT}    &{\tt =}& FORTRAN compiler options. \\
\end{tabular}

\item Compilation.

  By executing command {\tt make install} one executable, i.e.~the interface
  program({\tt grc4f}), is generated at {\tt BINDIR}.
  Furthermore three libraries, i.e.~{\tt BASES/SPRING}, {\tt CHANEL} and
  kinematics utility library, are generated in {\tt LIBDIR}.

\item Install default {\it Makefile}s.

  By executing {\tt src/lgen.sh} command,
  all {\it Makefile}s for 76 processes will be generated according
  to the environment where {\tt grc4f} has been installed.
\end{enumerate}

The sample control data files will be found at the directory
{\tt sample}.

%% file: testrun.tex
\leftline{\bf TEST RUN OUTPUT}

\leftline{\tt control data}

\begin{verbatim}
process = eNEuD
end
\end{verbatim}

Followings are the output files from {\tt BASES} and {\tt SPRING}.
Only one histogram, the energy distribution of the particle 1, is
shown since the whole output is too lengthy to be included here.
\newpage
\pagestyle{empty}
\centerline{\epsfbox{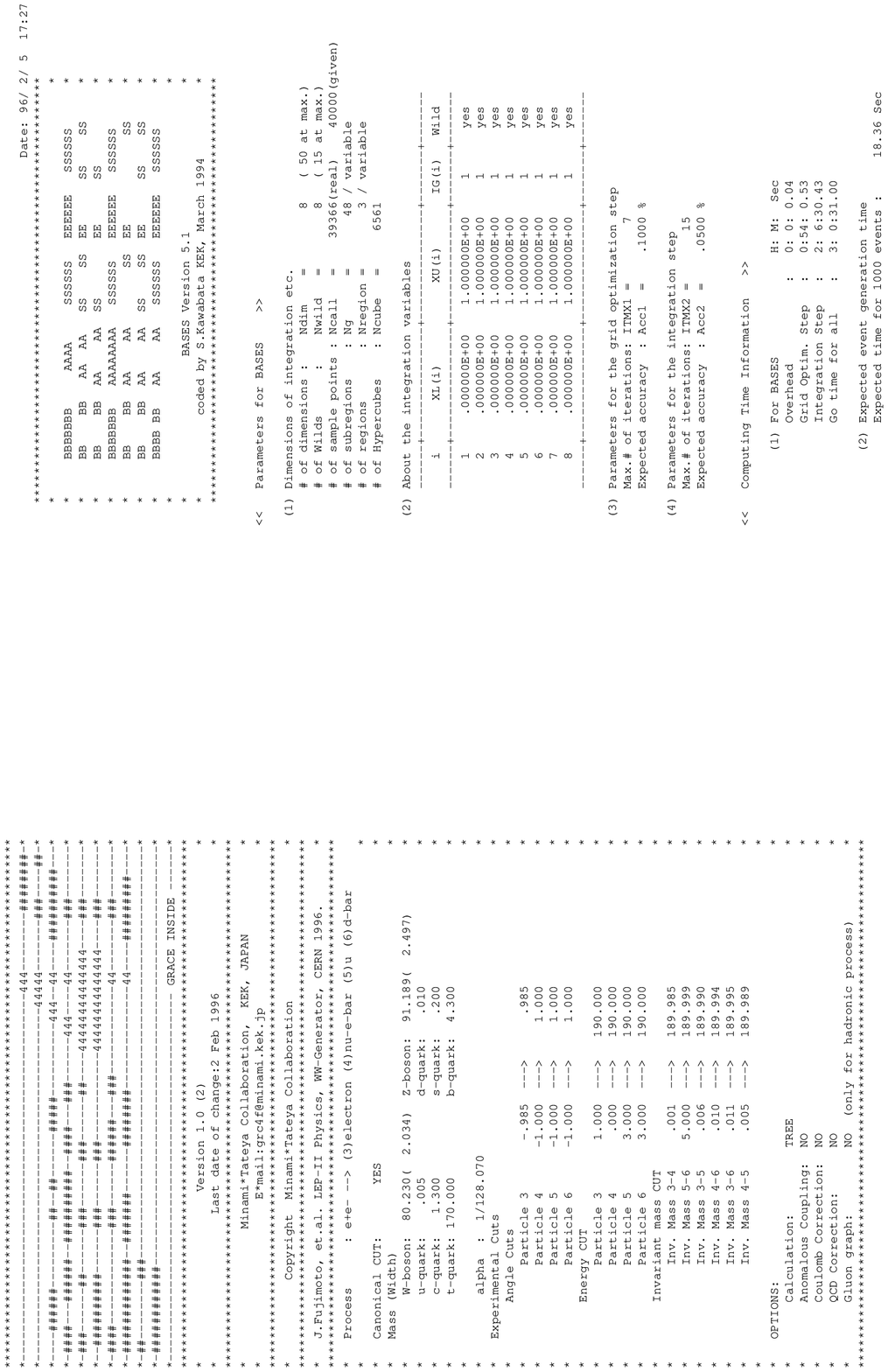}}
\centerline{\epsfbox{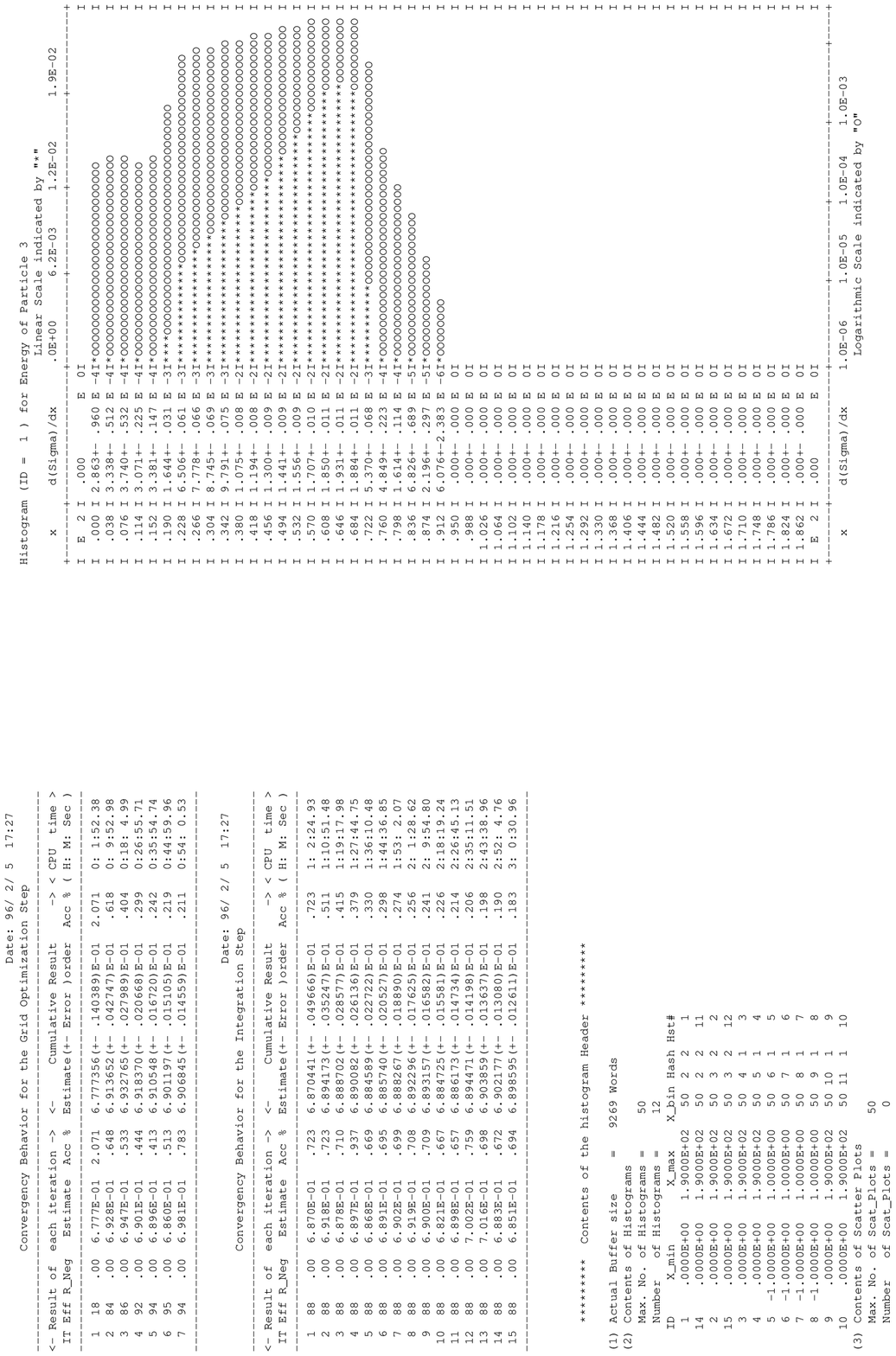}}

\centerline{\epsfbox{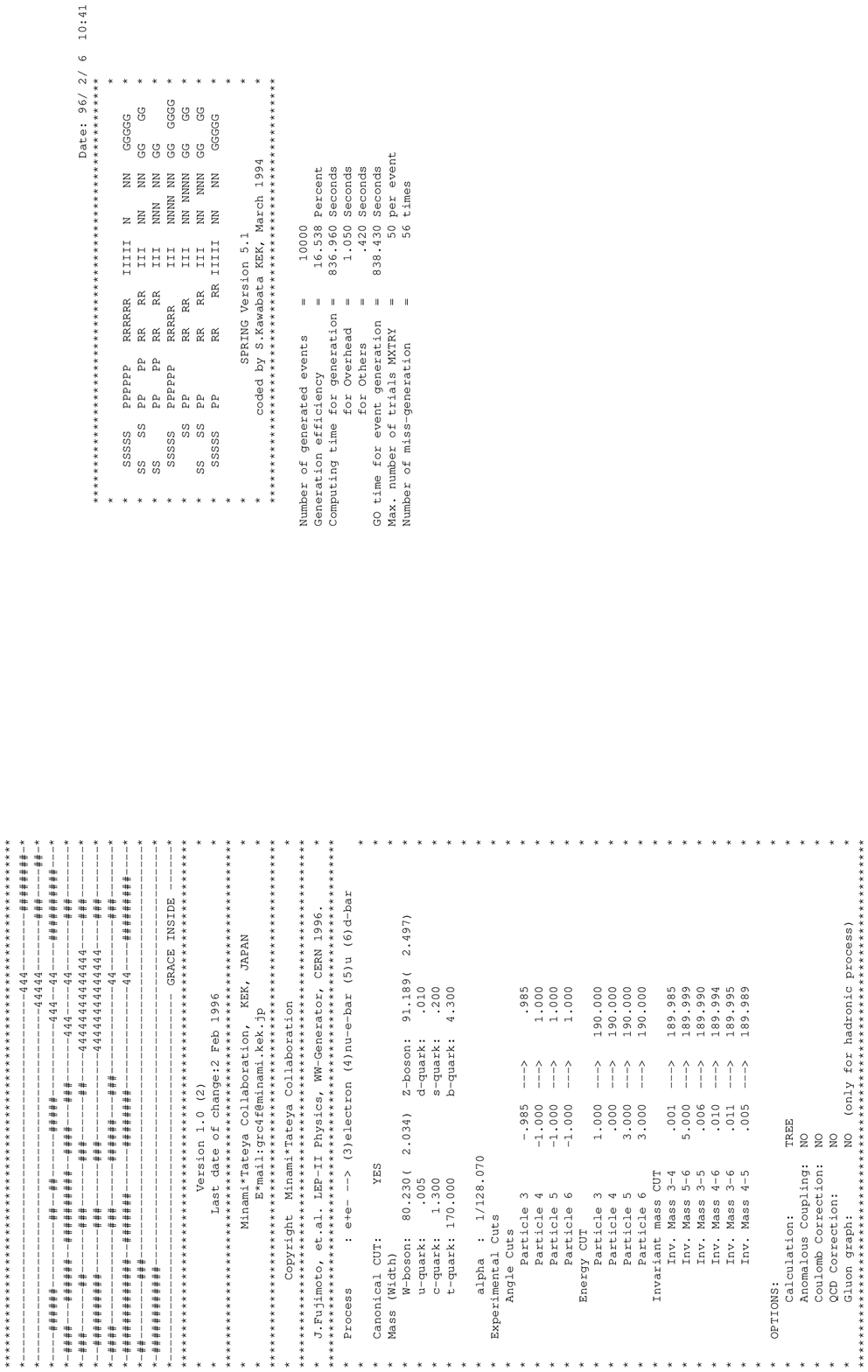}}
\centerline{\epsfbox{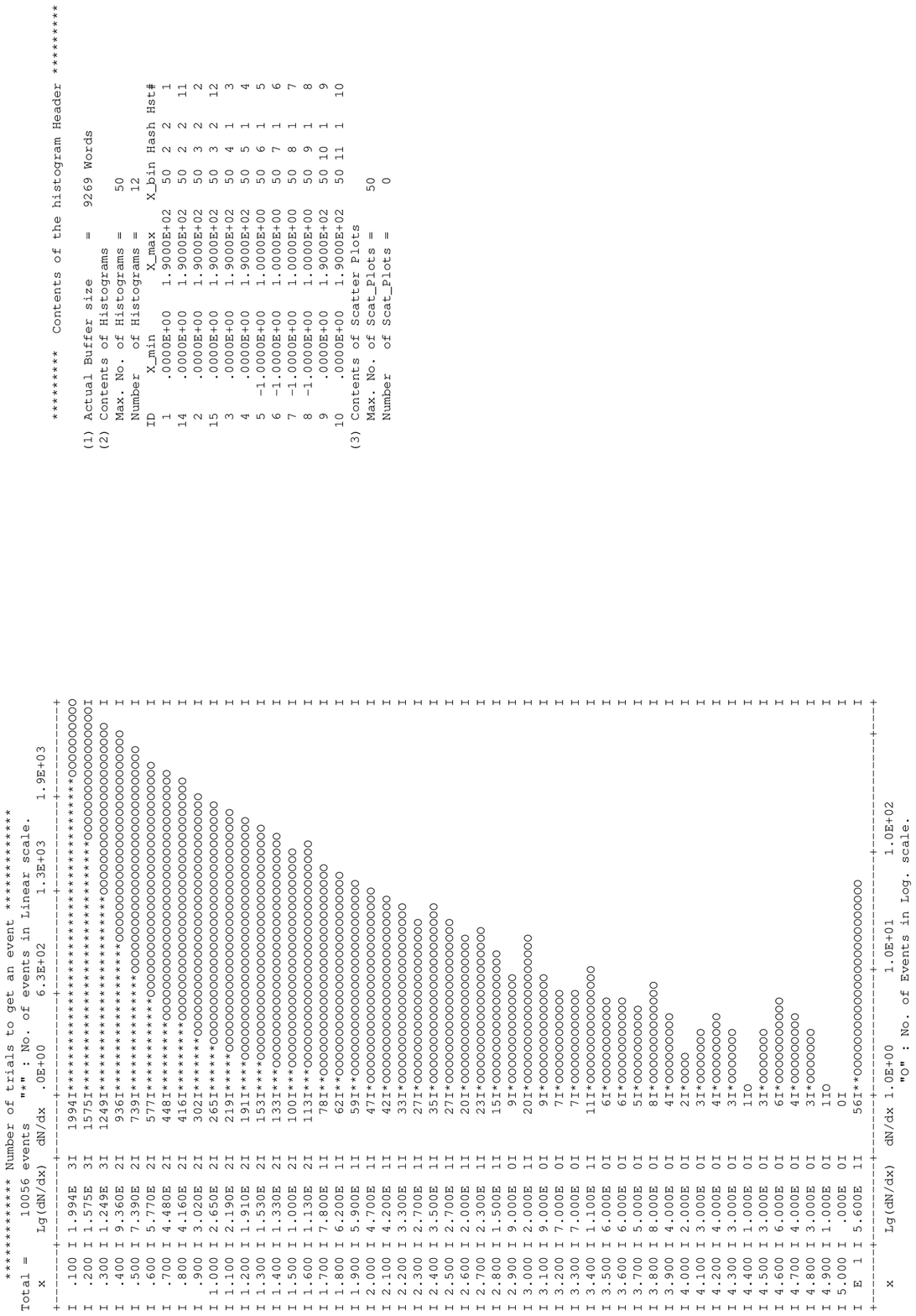}}
\centerline{\epsfbox{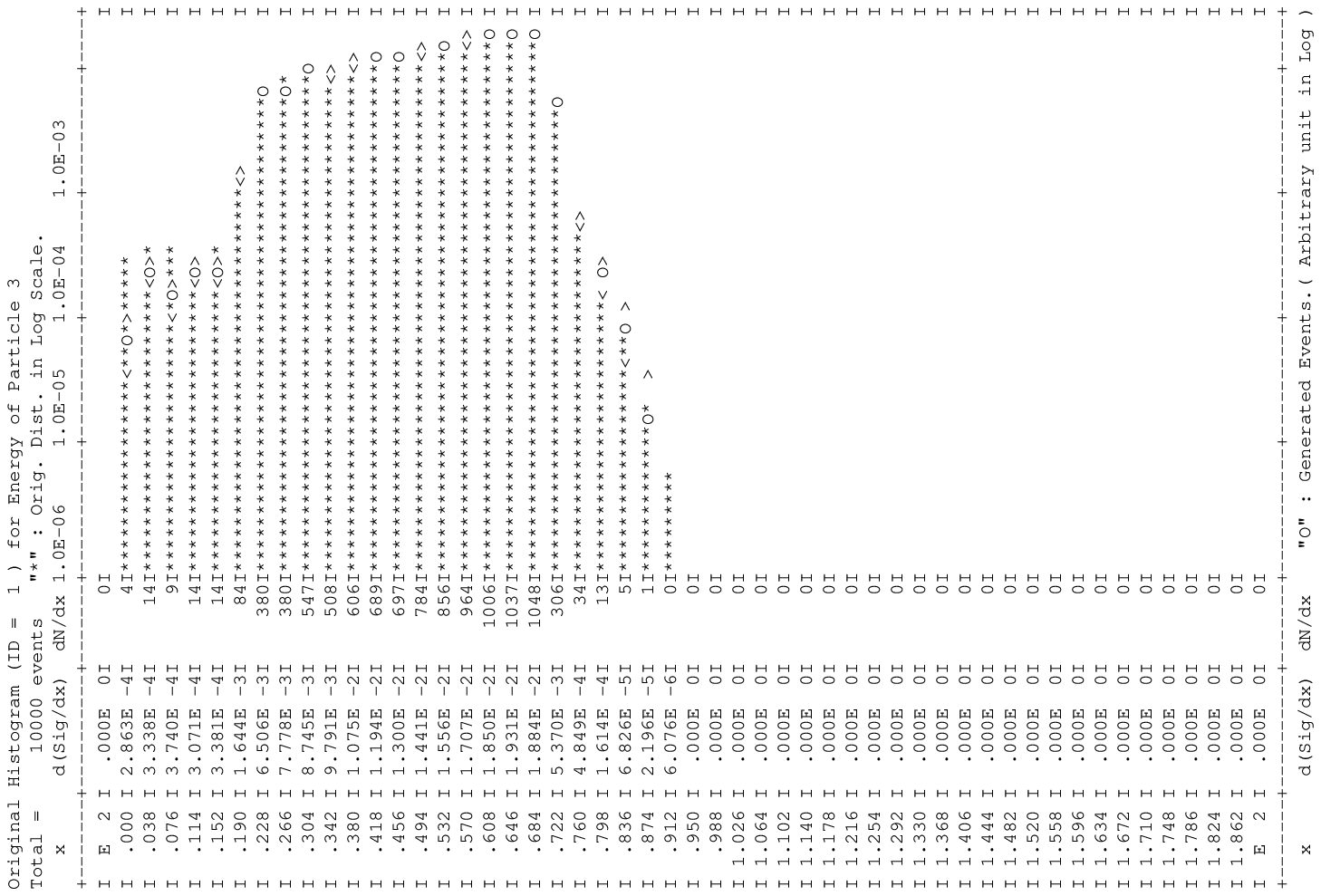}}
\newpage
\pagestyle{empty}